\documentstyle[preprint,isolatin1,eqsecnum,epsf,cite,aps]{revtex}
\begin{document}
\draft
\title{On the Nonrelativistic  Limit of the Scattering of Spin One-half 
Particles Interacting with a Chern--Simons Field}
\author{M. Gomes and A. J. da Silva}
\address{ Instituto de F\'\i sica, Universidade de S\~ao Paulo, CP66318,
05315--970, S\~ao Paulo, SP, Brazil.}
\date{October 2, 1997}

\maketitle 
\begin{abstract}
  Starting from a relativistic quantum field theory, we study the low
  energy scattering of two fermions of opposite spins interacting
  through a Chern-Simons field. Using the Coulomb gauge we implement
  the one loop renormalization program and discuss vacuum polarization
  and magnetic moment effects. We prove that the induced magnetic
  moments for spin up and spin down fermions are the same. Next, using
  an intermediary auxiliary cutoff the scattering amplitude is
  computed up to one loop. Similarly to Aharonov-Bohm effect for spin
  zero particles, the low energy part of the amplitude contains a
  logarithmic divergence in the limit of very high intermediary
  cutoff. In our approach however the needed counterterm is
  automatically provided without any additional hypothesis.
\end{abstract}

\section{Introduction} 
Recent studies have unveiled interesting aspects of the
scattering of charged particles by a thin magnetic flux tube, the
Aharonov-Bohm (AB) effect \cite{Ah}. Such investigations were
motivated by some conceptual difficulties in the case of spin zero
particles \cite{Fe}. In that situation, it has been shown that
accordance between the exact and perturbative calculations can be
achieved only after the inclusion in the perturbative method of a
contact, delta like interaction. As remarked in \cite{Lo}, the
discrepancy between the two results, prior to the addition of the
delta interaction, was to be expected since different boundary
conditions were adopted.  In a field theory context, taking the AB and
anyon scatterings in (2+1) dimensions as equivalent, the contact
interaction may be simulated by a $(\phi^* \phi)^2$, where $\phi$ is
the particle's field. This procedure was employed to reproduce the AB scattering
in a direct nonrelativistic approach \cite{Lo} and also as the low
energy limit of the relativistic theory of a scalar field minimally
coupled to a Chern--Simons field \cite{Bo}.

It has been argued that in the fermionic case the additional
auto--interaction is not needed since it is automatically provided by
Pauli's magnetic term.  These issues have been recently examined
\cite{Go,Ha} from two different perspectives.  In \cite{Go} we
examined the scattering of two low energy spin up fermions, starting
from a relativistic quantum field formulation and in \cite{Ha} the
same amplitude was calculated through a Galilean formulation of field
theory as described by L\'evy--Leblond \cite{Le}.  As expected, the
amplitudes obtained by both methods were free from ultraviolet
divergences, in accord with the above mentioned conjecture for the
scattering of two spin up fermions. Nevertheless, our results show
differences with respect to the nonrelativistic treatment coming from
vacuum polarization, induced magnetic moment and also from the
exchange of two ``photons''. In reference \cite{Ha}, on the other
hand, it was also considered the scattering of a spin up and a spin
down fermions, with the conclusion that the Pauli magnetic interaction
cancels and a logarithmic divergence shows up. Motivated by these
developments we here extend our study of \cite{Go} to the scattering
of a spin up and a spin down fermions starting from a relativistic
quantum field formulation. By introducing an intermediary cutoff,
we separate the regions of low and high virtual momenta in the Feynman
integrals. This allows a direct simplification of the integrands and
suggests the identification of the low momentum part with the corresponding
Galilean amplitude got in \cite{Ha}.  In our approach, however, the
logarithmic divergence is canceled by a contribution coming from the
complementary high momentum part, leaving us with a finite result.
We also determine an effective Lagrangian which summarizes the low momentum
behavior of a system of two relativistic fermions interacting through
a CS field.

\section {Tree approximation and one loop renormalization}

The Lagrangian of the system is
\begin{equation}
\label{1}
{\cal L} = \frac{\theta}{4}
\epsilon^{\mu\nu\alpha}F_{\mu\nu}A_{\alpha}
+\bar{\psi}(i \not \! \partial-m) \psi
+\bar{\varphi}(i \not \! \partial+m)\varphi
 +e \bar \psi\gamma^\mu \psi A_\mu+e \bar \varphi\gamma^\mu \varphi A_\mu,
\end{equation}
\noindent
where $F_{\mu\nu} = \partial_\mu A_\nu - \partial_\nu A_\mu$ and
$\psi$ ($\varphi$) is a two component Dirac field representing
particles and anti-particles of spin up (down) and the same mass $m$
(the parameter m is to be taken positive). We shall be interested in
getting the low energy approximation for the scattering amplitude of
one particle spin up by another of spin down (total spin zero). For completeness, we shall also reproduce the corresponding results for the case of two
spin up fermions but before anti-symmetrization, as done in \cite{Go}.
Assuming that in the center of mass frame the incoming and outgoing
particles have momenta $p_1=(w_p,\vec p)$, $p_2= (w_p,-\vec p)$ and
$p_{1}^{\phantom{x}\prime}= (w_p,\vec p^{\phantom{x} \prime})$,
$p_{2}^{\phantom{x}\prime}= (w_p,-\vec p^{\phantom{x} \prime})$, where
$|\vec p|= |{\vec p}^{\phantom {x}\prime}|$ and $w_p= \sqrt{m^2+ {\vec
    p}^2}$, the tree approximation for these processes are
respectively \cite{o1},
\begin{eqnarray}
\label{3}
T^{(0)}_{\uparrow\downarrow}& = & -i e^2 \bar v(\vec {p}^{\phantom {x}\prime})\gamma^\mu v(\vec {p})
\Delta_{\mu\nu}(\vec {p}^{\phantom {x}\prime}-\vec {p}) \bar 
u(-\vec {p}^{\phantom {x}\prime})\gamma^\nu u(-\vec {p})\\
\label{2}
T^{(0)}_{\uparrow\uparrow}& = &-i e^2 \bar u(\vec {p}^{\phantom {x}\prime})\gamma^\mu u(\vec {p})
\Delta_{\mu\nu}(\vec {p}^{\phantom {x}\prime}-\vec {p}) \bar 
u(-\vec {p}^{\phantom {x}\prime})\gamma^\nu u(-\vec {p})
\end{eqnarray}
where the CS field propagator in the Coulomb gauge \cite{Ko1} is given by

\begin{equation}
\Delta_{\mu\nu}(k)= \frac{1}{\theta}\epsilon_{\mu\nu\rho}
\frac{\bar k^\rho}{\vec k^2},\label{4}
\end{equation}

\noindent
with $\bar k^\alpha \equiv (0,\vec k)$. 

After expanding in powers of $\frac{|\vec p|}{m}\; (\ll 1)$, we get in leading
order,

\begin{eqnarray}
 \label{6}
T^{(0)}_{\uparrow\downarrow}& = &\frac {i e^2 }{\theta m}\frac{\vec s \wedge\vec q}{{\vec q}^2}\\
\label{5}
T^{(0)}_{\uparrow\uparrow}& = &\frac {e^2}{\theta m} (1 + i\frac{\vec s \wedge \vec q}{{\vec q}^2})
\end{eqnarray}
 
\noindent
where $\vec s=\vec p + \vec p^{\phantom{x}\prime}$, $\vec q = \vec
p^{\phantom{x}\prime}-\vec p$ and $\vec s \times \vec q$ stands for
$\epsilon^{ij}s_i q_j$.  As can be seen in a direct nonrelativistic
treatment \cite{Go,Ha}, the origin of the constant term in Eq.
(\ref{5}) is a contact Pauli interaction between the magnetic moment
of each fermion with the magnetic field of the other. In the
anti-parallel case, Eq. (\ref{6}), these effects cancel each other and
this is the basic reason behind the above mentioned discrepancy for
the spin zero scalar case. 

Before embarking into the discussion of the one loop scattering amplitudes, let
us comment on the renormalization of the model (\ref{1}). As in the case of
just one fermion field interacting with the CS field, \cite{Gi}, the divergences are
concentrated into contributions to the vacuum polarization, self energies and vertex parts.
The analysis is entirely analogous to that of Ref. \cite{Gi}. Now, however,there are two contributions to the CS self energy,

\begin{equation}\label{7}
\Pi^{\alpha \beta}(k)\,= \, -e^2\int\frac{d^3q}{(2\pi)^3}\frac{{\rm
Tr}[\gamma^\alpha(\not \!q+\not \!k+m)\gamma^\beta(\not \!q
+m)]}{[(k+q)^2-m^2](q^2-m^2)} + (m\rightarrow -m) ,
\end{equation}

\noindent
and the would be induced CS term vanishes. Therefore $\theta$ is kept unchanged and can
be considered as the renormalized CS parameter. For low momentum we get

\begin{equation}
\label{8}
\Pi^{\alpha\beta}(k)= - i\frac {e^2}{6 \pi m}(k^2 g^{\alpha\beta}- k^\alpha 
k^\beta), 
\end{equation}
\noindent
implying that the effective low energy Lagrangian will contain a
Maxwell term with intensity twice as big as in the case of just one
fermion field.

Defining the $\psi$ field self-energy, $\Sigma_{\psi}$ from
the propagator $S_{\psi F}(p)=i(\not \!p -m +i \Sigma_{\psi})^{-1}$, we have

\begin{equation}\label{9}
 \Sigma_{\psi}(p) = -\frac {i e^2 }{\theta} \int \frac {d^3k}{(2 \pi)^3}
\frac{\gamma^\mu (\not \!k+ \not \!p +m)\gamma^\nu}{(k+p)^2-m^2}
\frac {\epsilon_{\mu\nu\rho}{\bar k}^{\rho}}{({\vec k})^2}\, .
\end{equation}

Choosing a mass counterterm so that $m$ is the physical mass, i. e., the
position of the propagator's pole, we obtain the renormalized result
\begin{equation}
\Sigma_{\psi}=- \frac {i e^2}{ 2 \pi \theta}\Bigl [\frac {({\vec p}^2- m 
\vec p \cdot \vec \gamma)}{m+w_p}  \Bigr ].\label{10a}
\end{equation}

\noindent
The self--energy of the $\varphi$ field is very
similar, giving ($S_{\varphi F}(p)=i(\not \!p +m +i \Sigma_{\varphi})^{-1}$)
\begin{equation}
\Sigma_{\varphi}=- \frac {i e^2}{ 2 \pi \theta}\Bigl [\frac {({\vec p}^2+ m 
\vec p \cdot \vec \gamma)}{m+w_p}  \Bigr ]\label{10}.
\end{equation}

There are two vertex parts to be considered. The calculation of these vertex parts is  greatly simplified if it is restricted to the low momentum region.  For the $\varphi$ field up to one loop we have, 
\begin{equation}
\label{11}
\Gamma^{\rho}_{\varphi}(p, p') =
\frac {i e^3}{\theta} \int \frac {d^3k}{(2 \pi)^3}
 \frac{\gamma^\mu (\not \!p' - \not \!k-m)\gamma^\rho (\not \!p - 
\not \!k-m)\gamma^\alpha \epsilon_{\alpha \mu \nu}{\bar k}^\nu }{[(p'-k)^2 
- m^2+i \epsilon][(p-k)^2 - m^2+i \epsilon](-{\vec k}^2)}+ i e \gamma^\rho
\end{equation}  

\noindent
so that, in the low momentum regime,
\begin{eqnarray}
\label{12a}
\bar{v}(p') \Gamma_{\varphi}^{0} v(p) &=& ie, \qquad \\
\bar{v}(p')\Gamma_{\varphi}^{i} v(p) &=& i e\left (1 + 
\frac{e^2}{4\pi\theta}\right ) \frac{(p+p'^i)}{2 m} - e \left(
1-\frac{e^2}{4 \pi \theta}\right ) \epsilon^{ij}\frac{(p'-p)^j}{2m}.\end{eqnarray}

The vertex part of the $\psi$ field is  similar
and one only has to observe the changes due to the different sign of the mass (see Ref. \cite{Go})

Coupling, alternatively, an external electromagnetic field to the
fields $\varphi$ and $\psi$, one could compute quantum corrections to
their respective magnetic moments, Fig. 1. We would like to emphasize
that, contrarily to what happens in covariant gauges, in the Coulomb
gauge it is essential to take into account self energy corrections to
obtain the correct result.  In particular, for spin down fermions
scattered by an effective external field ${\cal A}^\mu (q)$, which
already incorporates the polarization of the vacuum, we obtain the
scattering amplitude
\begin{equation}\label{12b}
i e {\cal A}^0(q) - i \frac {e}{2m}  {s^i} {\cal A}^i(q) + \frac{e}{2m} (1-\frac{e^2}{2 \pi \theta}) \epsilon^{ij}q^j {\cal A}^i(q).
\end{equation}

From this expression, one sees that the sole effect of the vertex correction
is to modify the fermion magnetic moment which now turns out to be
\begin{equation}\label{12c}
\mu_{down} = -\frac e{2m}\Bigl ( 1 - \frac {e^2}{4 \pi \theta}\Bigr).
\end{equation}
This result must be compared with 
\begin{equation}\label{12d}
\mu_{up} = \frac e{2m}\Bigl ( 1 + \frac {e^2}{4 \pi \theta}\Bigr )
\end{equation}
got in \cite{Go} for the spin up field $\psi$. It should be noticed
that the induced magnetic moment has the same sign in both cases.
This is in accord with the results for the anomalous spin
\cite{Ha,Si,Ko}.  Therefore, the total intensity of the magnetic
moment increases for one field and lowers for the other.

\section{One loop scattering}
We are now ready to pursue our analysis of the scattering of spin 1/2
particles at one loop order. First of all, self--energy and radiative
corrections to the tree approximation, in leading $1/m$ order, give
\begin{equation}\label{12e}
T_{\uparrow\downarrow R}= \frac{ e^4}{6 \pi m\theta^2}+\frac{ e^4}{2 
\pi m\theta^2}= \frac{2 e^4}{3 \pi m\theta^2},
\end{equation}
where the first and second terms in the first equality come,
respectively, from the vacuum polarization and vertex insertions. It is
easy to verify that the same expression holds for the spin-up/spin-up
case ( the difference of this result with respect to the one in
\cite{Go} is just because we now have two fields instead of one
contributing to the vacuum polarization tensor).

The remaining graphs to be analyzed are shown in Fig. 2.
The box and crisscross two photon exchange
amplitudes, for the total spin zero case, are given by 
\begin{eqnarray}
\label{13}
&& T_{\uparrow\downarrow B} =i e^4 \int \frac {d^3 k}{(2 \pi)^3} \Bigl [\bar 
v(\vec {p}^{\phantom {x}\prime})\gamma^\mu S_F(r,-m) \gamma^\nu 
v(\vec p)\Bigr ] 
\nonumber \\
&&\Bigl [\bar u(-\vec {p}^{\phantom {x}\prime})\gamma^\alpha S_F(r^\prime, m) 
\gamma^\beta u(-\vec p)\Bigr ] \Delta_{\nu\beta}(\vec k - \vec p) 
\Delta_{\alpha\mu}(\vec k - \vec p^{\phantom {x}\prime})
\end{eqnarray}
and
\begin{eqnarray}
\label{14}
&& T_{\uparrow\downarrow X}=i e^4 \int \frac {d^3 k}{(2 \pi)^3} \Bigl [\bar 
v(\vec {p}^{\phantom {x}\prime})\gamma^\mu S_F(r,-m) \gamma^\nu 
v(\vec p)\Bigr ] 
\nonumber \\
&&\Bigl [\bar u(-\vec {p}^{\phantom {x}\prime})\gamma^\alpha S_F(t,m) 
\gamma^\beta u(-\vec p)\Bigr ] \Delta_{\nu\alpha}(\vec k - \vec p) 
\Delta_{\beta\mu}(\vec k - \vec p^{\phantom {x}\prime}).
\end{eqnarray}

\noindent
where $r\equiv (w_p+k^0, \vec k )$, $r^\prime\equiv (w_p -k^0, - \vec k)$ and
$t \equiv (w_p +k^0, \vec k - \vec p- \vec p^{\phantom {x}\prime})$.

To simplify
the calculation we shall use
\begin{equation}
S_F(p, -m)=i \frac{\not \!p -m}{p^2 -m^2 + i \epsilon}=
i\frac{v(\vec p)\bar v(\vec p)}{p^0-w_p +i\epsilon}+i\frac{u(-\vec p)\bar
u(-\vec p)}{p^0+w_p -i\epsilon}.\label{15}
\end{equation}
for the $\varphi$ free field propagator. The propagator of the free $\psi$ field,
$S_F(p,m)$,
is obtained from (\ref{15}) replacing $m$ by $-m$ and exchanging the spinors
$u$ and $v$. Using these expressions we get, 

\begin{eqnarray}\label{16}
&& T_{\uparrow\downarrow B}=- ie^4 \int \frac {d^3 k}{(2 \pi)^3} \Delta_{\nu\beta}(
\vec k - \vec p) \Delta_{\alpha\mu}(\vec k - \vec p^{\phantom {x}\prime}) 
\nonumber \\
&&\bar v (\vec p^{\phantom {x}\prime}) \gamma^\mu \Bigl [ \frac {v(\vec k) 
\bar v (\vec k)}{k^0+ w_p-w_k + i \epsilon}+ \frac {u(-\vec k)
\bar u(-\vec k)}{k^0+ w_p+w_k - i \epsilon}\Bigr ]\gamma^\nu 
v (\vec p)\nonumber \\
&& \bar u (-\vec p^{\phantom {x}\prime}) \gamma^\alpha \Bigl 
[ \frac {u(-\vec k) \bar u (-\vec k)}{w_p-k^0-w_k + i \epsilon}+ 
\frac {v(\vec k) \bar v(\vec k)}{w_p-k^0+w_k - i \epsilon}\Bigr ]
\gamma^\beta u (-\vec p)
\end{eqnarray}

and

\begin{eqnarray}\label{17}
&& T_{\uparrow\downarrow X}=-i e^4 \int \frac {d^3 k}{(2 \pi)^3} \Delta_{\nu\alpha}(
\vec k - \vec p) \Delta_{\beta\mu}(\vec k - \vec p^{\phantom {x}\prime}) 
\nonumber \\
&&\bar v (\vec p^{\phantom {x}\prime}) \gamma^\mu \Bigl [ \frac {v(\vec k) 
\bar v (\vec k)}{k^0+ w_p-w_k + i \epsilon}+ \frac {u(-\vec k)\bar 
u(-\vec k)}{k^0+ w_p+w_k - i \epsilon}\Bigr ]\gamma^\nu v (\vec p)
\nonumber \\
&& \bar u (-\vec p^{\phantom {x}\prime}) \gamma^\alpha \Bigl [ 
\frac {u(\vec k- \vec s) \bar u (\vec k-s)}{k^0+w_p-w_{k-s} + i \epsilon}
+ \frac {v(\vec s-\vec k) \bar v(\vec s-\vec k)}{k^0+w_p+w_{k-s} 
- i \epsilon}\Bigr ]\gamma^\beta u (-\vec p).
\end{eqnarray}

After integrating in $k^0$ and some simplifications, we obtain
\begin{equation}\label{18}
T_{\uparrow\downarrow B} = T_{Bvu}+T_{Buv}
\end{equation}

\noindent
where
\begin{eqnarray}\label{19}
T_{Bvu}&=& - \frac{e^4}{2} \int \frac{d^2k}
{(2 \pi)^2}\frac {w_k+w_p}{\vec k^2-\vec p^2-i\epsilon} F^{*}
(\vec k,\vec p^{\phantom {x}\prime}) F(\vec k,\vec p),\\
T_{Buv}&=&- \frac{e^4}{2} \int \frac{d^2k}{(2 \pi)^2}
\frac{1}{w_k+w_p}H(\vec p^{\phantom {x}\prime},\vec k) H^*(\vec p,\vec k),
\end{eqnarray} 

\noindent
whereas for the crisscross graph it results,

\begin{eqnarray}\label{20}
T_{\uparrow\downarrow X} &=& \frac{e^4}{2} \int \frac{d^2k}{(2 \pi)^2}\frac {1}
{w_k+w_{k-s}}\Bigl[ G_1(\vec k -\vec s,-\vec p,\vec p^{\phantom {x}\prime}) 
G_{1}^{*}(\vec k -\vec s,-\vec p^{\phantom {x}\prime},\vec p)\Bigr.\nonumber\\
&\phantom{a}&  \Bigl.+  G_2(\vec k,\vec p,-\vec p^{\phantom {x}\prime}) G_{2}^{*}(\vec k,\vec p^{\phantom {x}\prime},-\vec p)\Bigr].
\end{eqnarray}

In the above formula $\vec s = \vec p + \vec p^{\phantom{x}\prime}$ and
\begin{eqnarray}
\label{21}
 F(\vec k, \vec p) &=& [\bar v(\vec {k})\gamma^\mu v(\vec {p})]
\Delta_{\mu\nu}(\vec {k}-\vec {p})[ \bar u(-\vec {k})\gamma^\nu 
u(-\vec {p})],\nonumber\\
 H(\vec p, \vec k) &=&  [\bar v(\vec {p})\gamma^\mu u(-\vec {k})]
\Delta_{\mu\nu}(\vec {p}-\vec {k}) [\bar u(-\vec {p})\gamma^\nu 
v(\vec {k})],\\
 G_1(\vec a, \vec b,\vec c)& = &[ \bar u(\vec {a})\gamma^\mu u(\vec {b})]
\Delta_{\mu\nu}(\vec {a}-\vec {b})[ \bar v(\vec {c})\gamma^\nu 
u(\vec b-\vec {a}-\vec c)]\nonumber\\
G_2(\vec a, \vec b,\vec c)& = &[ \bar v(\vec {a})\gamma^\mu v(\vec {b})]
\Delta_{\mu\nu}(\vec {a}-\vec {b})[ \bar u(\vec {c})\gamma^\nu 
v(\vec b-\vec {a}-\vec c)].\nonumber
\end{eqnarray}

To perform the spatial integration, we follow a scheme explained in
detail for the $\lambda \phi^4$ scalar theory in \cite{Ma}. We
separate the integration region into two parts through the
introduction of an auxiliary cutoff $|\vec p| \ll \Lambda \ll m$. In
the {\it low} energy part, $|\vec k| \leq \Lambda$, nonrelativistic
approximations are done directly in the integrands. In the
complementary {\it high} energy region, $|\vec k| \geq \Lambda$, we
Taylor expand the integrands around zero external momenta.  
This procedure is closely related to the methods of effective field
theories \cite{Wi}.
With these
simplifications, the above expressions become,

\begin{eqnarray}
T_{Bvu} &=& - \frac{4e^4}{\theta^2 m}\int_{0}^{\Lambda}
\frac{d^2k}{(2\pi)^2}\; \frac{1}{{\vec k}^2 -{\vec p}^2 -i\epsilon}
\frac{{\vec k} \wedge {\vec p}}{({\vec k}-{\vec p})^2}
\frac{{\vec k} \wedge {\vec p^{\phantom {x}\prime}}}{({\vec k}
-{\vec p^{\phantom {x}\prime}})^2}\nonumber \\
&\phantom {x}&  -\frac{e^4}{2m^2 \theta ^2} 
 \int _{\Lambda}^{\infty} \frac{d^2k}{(2\pi)^2}\; \frac{(w_k +m)^3}{{|\vec k|}^2 w_k^2} \frac{{\vec k} \wedge {\vec p}}{|\vec k|^2} \;\frac{{\vec k} \wedge {\vec p^{\phantom {x}\prime}}}{|\vec k|^2},
\label{22}\\
T_{Buv} &=&-\frac{e^4}{16\theta^2 m^7} \int_{0}^{\Lambda}
\frac{d^2k}{(2\pi)^2}\;({\vec k} \wedge {\vec p}) ({\vec k} \wedge {\vec p^{\phantom {x}\prime}})- \frac{e^4}{2\theta^2 m^2}\int _{\Lambda}^{\infty} \frac{d^2k}{(2\pi)^2}\;\frac{({\vec k} \wedge {\vec p}) ({\vec k} \wedge {\vec p^{\phantom {x}\prime}})}{w_{k}^{2}(m+w_k)^3 }
\end{eqnarray}

and
\begin{equation}\label{23}
T_{\uparrow\downarrow X} = \frac{e^4}{m \theta^2} \int_{0}^{\Lambda} \frac{d^2k}{(2 \pi)^2}
\frac{(k-p)_{-}}{({\vec k}-{\vec p})^2} \, \frac{(k-p^\prime)_{+}}{({\vec k}-\vec p^{\phantom {x}\prime})^2}+\frac{e^4}{\theta^2}\int _{\Lambda}^{\infty}
\frac{d^2k}{(2\pi)^2} \frac{1}{w_k \vec k^2},
\end{equation}
where $a_{-}= a^1 - i a^2$ and $a_{+}= a^1 +i a^2$. Performing the
integrals and keeping only the dominant terms in $1/m$, we find,
\begin{eqnarray} \label{24}
T_{Bvu}&=& \Bigl[\frac{e^4}{4 \pi m \theta^2} \ln\Bigl (- \frac{q^2}{p^2 + i \epsilon} \Bigr )\Bigr ]_{low}\\
T_{\uparrow\downarrow X}&=&\Bigl[\frac{e^4}{4 \pi m \theta^2} \ln\frac{\Lambda^2}{q^2}\Bigr ]_{low}
+ \Bigl [\frac{e^4}{4 \pi m \theta^2} \ln \frac {4 m^2}{\Lambda^2}
\Bigr ]_{high}.
\label{25}
\end{eqnarray}
We observe that, up to leading order, $T_{Buv}$ and the {\it high} part of
$T_{Bvu}$ vanish. 
Differently from what happens in the Galilean formulation \cite{Ha}, our
result is finite,
\begin{equation}\label{25a}
T_{\uparrow\downarrow B}+T_{\uparrow\downarrow X}=  \frac{e^4}{4 \pi m \theta^2} \ln\Bigl (- \frac{4 m^2}{p^2 + i \epsilon} \Bigr ).
\end{equation}
Nevertheless, it should be noticed that
if we consider only the low energy contributions and reinterpret $\Lambda$
as an ultraviolet cutoff then the scattering amplitude for the crisscross and box diagrams diverges logarithmically,i. e.,
\begin{equation} \label{Ha1}
[T_{\uparrow\downarrow B}+T_{\uparrow\downarrow X}]_{low}= \frac{e^4}{4 \pi m \theta^2} \ln\Bigl (- \frac{\Lambda^2}{p^2 + i \epsilon} \Bigr ).
\end{equation}
 This is exactly what happens in the nonrelativistic theory
were the amplitude only becomes finite after the addition of a quartic
counterterm (Eq. (\ref{Ha1}), up to an overall phase, agrees with the last formula of \cite{Ha} for
$ss'=-1$ and $M=M'=m$, after the identification $g^2= e^2/\theta$). In our formulation however the needed counterterm is automatically provided
by the contribution from the {\it high} energy part of the above integrals.
In fact, in the nonrelativistic formulation the last term in (\ref{25}) can be effectively interpreted as coming
from the counterterm 
\begin{equation} \label{25b}
\frac{e^4}{2 \pi m \theta^2} \ln \frac {\Lambda}{2 m} (\bar \varphi \varphi)
(\bar \psi \psi),
\end{equation}
which suggests that a $(\bar \varphi \varphi) (\bar \psi \psi)$ interaction
should be considered from the starting in the nonrelativistic model.
The extra coupling parameter associated with such interaction could be
adjusted to reproduce up to this order the expansion of the
non-perturbative AB effect, canceling the one-loop contributions.

The whole scattering one loop amplitude of one spin up and one spin down
fermions is given by the sum of Eq. (\ref{12e}) and (\ref{25a}) 
\begin{equation}\label{25c}
T_{\uparrow\downarrow R}+T_{\uparrow\downarrow B}+T_{\uparrow\downarrow X}=
\frac{2 e^4}{3 \pi m \theta^2}+  
\frac{e^4}{4 \pi m \theta^2} \ln\Bigl (- \frac{4 m^2}{p^2 + i \epsilon} \Bigr )
\end{equation}
and it
is finite. The corresponding result for the scattering of two spin up
fermions was calculated in \cite{Go}.  Taking into consideration the
modification in the vacuum polarization due to the presence of the
spin down fermion one obtains the following one loop amplitude for two
spin up fermions before
anti-symmetrization,
\begin{equation}\label{26}
T_{\uparrow \uparrow R}+ T_{\uparrow \uparrow B}+T_{\uparrow \uparrow X} = 
\frac{e^4}{6 \pi m \theta^2}.
\end{equation}
As remarked in \cite{Go}, in the case of just one flavor this
contribution actually vanishes after anti-symmetrization. However, in Eq. (\ref{25c}) the two fermions are distinguishable and no anti-symmetrization is
required; it survives as it stands. 

The effective low energy Lagrangian emerging from our study is given by
\begin{eqnarray}
{\cal L}_{eff} &=& \psi^\dagger (i\frac{d\phantom{x}}{dt}- e A^0) \psi
- \frac{1}{2 m} |\vec \nabla\psi - i e \vec A \psi)|^2 + e^2/2\pi\theta) \, 
B\psi^\dagger \psi \nonumber \\ &\phantom&  
 + \varphi^\dagger (i\frac{d\phantom{x}}{dt}- e A^0) \varphi
- \frac{1}{2 m} |\vec \nabla\varphi - i e \vec A \varphi|^2-\frac{e}{2m}\,(1
- e^2/2\pi\theta) \, B \varphi^\dagger \varphi \nonumber \\ &\phantom&  + 
\frac{\theta}{4} \epsilon_{\mu\nu\rho} A^\mu
F^{\nu\rho} - \frac{1}{4}\left (\frac{\, e^2}{6 \pi m}\right
)F^{\mu\nu}F_{\mu\nu} +\frac{e^4}{2 \pi m \theta^2} \ln \frac {\Lambda}{2 m} 
(\bar \varphi \varphi)(\bar \psi \psi).
\end{eqnarray}

Up to one loop this Lagrangian summarizes the low momentum behavior
of a system of two relativistic fermions interacting through a CS
field. It differs from the usual Pauli--Schr\"odinger Lagrangian for
nonrelativistic fermions minimally coupled to a CS field, due to the
incorporation of purely quantum high energy effects: anomalous
magnetic moments, vacuum polarization and of a quartic fermionic
counterterm. This last term results from the contribution of
relativistic intermediate states to the scattering amplitude of spin
up and spin down fermions. The parameter $\Lambda$ is to be understood as
an ultraviolet cutoff to be used in the computations with the above
effective Lagrangian.
\begin{center}
ACKNOWLEDGMENTS  
\end{center}

This work was supported in part by Conselho Nacional de
Desenvolvimento Cient\'\i fico e Tecnol\'ogico (CNPq) e Funda\c c\~ao de
Amparo \`a Pesquisa do Estado de S\~ao Paulo (FAPESP).

\begin{figure}
\centerline{\epsfbox{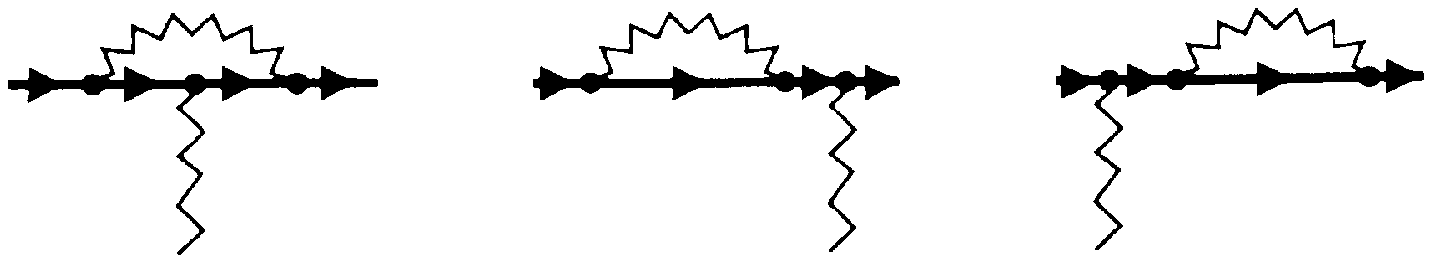}} 
\label{fig1} 
\end{figure}

FIG. 1. One loop contributions to the fermion anomalous magnetic moment.

\vspace{1.0cm}
\begin{figure} 
{\centerline{\epsfbox{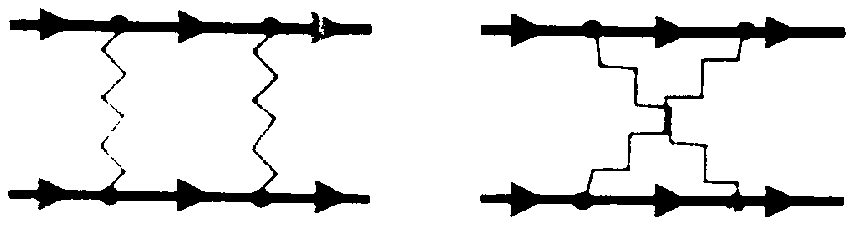}}} 
\label{fig2}
\end{figure}

FIG. 2. Graphs contributing to the relativistic fermion-fermion scattering in 
one loop approximation.

\begin{references}
\bibitem {o1} We use natural units $(c=\hbar=1)$ and our metric is
$g_{00}=-g_{11}=-g_{22}=1$. The fully antisymmetric tensor
$\epsilon^{\mu \nu \lambda}$ is normalized such that
$\epsilon^{012}=1$ and we define $\epsilon^{ij} \equiv
\epsilon^{0ij}$.  Repeated greek indices sum from 0 to 2, while
repeated Latin indices from the middle of the alphabet sum from 1 to
2.  For the $\gamma$-matrices we adopt the representation
$\gamma^{0}=\sigma^{3},\gamma^{1}=i\sigma^{1},\gamma^{2}=i\sigma^{2}$,
where $\sigma^{i}, i=1,2,3,$ are the Pauli spin matrices.  The
solutions of the free Dirac equation corresponding to particles of spin up
an down are given by

\[
u(\vec{p})\,\,=\,\, \left( \frac{w_p+m}{2\,w_p}
\right)^{\frac {1}
{2}}\,\left[ \begin{array}{c} 1 \\
\frac{p^{2}\,-\,ip^{1}}{w_p\,+\,m} \end{array} \right] \,\,\quad 
v(\vec{p})\,\,=\,\, \left( \frac{w_p+m}{2\,w_p}
\right)^{\frac {1}
{2}}\,\left[ \begin{array}{c} \frac{p^{2}+i p^{1}}{w_p\,+\,m} \\
1\end{array} \right] \,\,.
\]
where $w_p=(m^2+{\vec p}^2)^{1/2}$ and the normalizations were chosen so that
$\bar u u = - \bar v v= m/w_p$.
\end{references}
\end{document}